\documentclass[apj]{emulateapj}

\shortauthors{COWIE, HU \& SONGAILA}
\shorttitle{LY$\alpha$-SELECTED GALAXIES AT $z=6.5$}
\begin{document}
\singlespace
\title{The UV-Continuum properties of Ly$\alpha$-Selected Galaxies at $z=6.5$}
\author{
Lennox L. Cowie\altaffilmark{1}\email{cowie@ifa.hawaii.edu}
Esther M. Hu\altaffilmark{1}\email{cowie@ifa.hawaii.edu}
Antoinette Songaila\altaffilmark{1}\email{acowie@ifa.hawaii.edu}
}

\affil{Institute for Astronomy, University of Hawaii,
  2680 Woodlawn Drive, Honolulu, HI 96822}

\altaffiltext{1}{Visiting Astronomer, W. M. Keck Observatory, which is jointly
  operated by the California Institute of Technology, the University of
  California, and the National Aeronautics and Space Administration}

\slugcomment{To be published in {\it Astrophysical Journal Letters}}

\begin{abstract}
We report the first space-based very deep near-infrared continuum observations 
of a uniform sample of $z=6.5$\ galaxies with log(L(L$\alpha$))$=42.5-43$ erg s$^{-1}$\
selected from narrow-band line searches
and with spectroscopically confirmed 
Ly$\alpha$\ emission. The 1.4$\mu$m HST WFC3 observations are
deep enough (AB($1\sigma) = 28.75$) to measure individual continuum magnitudes
at this redshift for all of the objects. 
We compare the results with continuum-selected samples at the same redshift and 
find that Ly$\alpha$\ emission is present in $24\%$ 
of all galaxies with $M_{\rm AB}(1350\rm\AA) < -20$\  at $z = 6.5$. 
The error in this quantity is dominated by systematic 
uncertainties, which could be as large as multiplicative factors of three.  
The Ly$\alpha$\ galaxies are extended but small (size $< 1~{\rm kpc}$),
and have star formation rates of approximately $10~M_{\odot}\  {\rm yr}^{-1}$.  
We find a mean $L({\rm Ly}\alpha)/\nu L_{\nu}$\ at 1400~\AA\  of 0.08, with the seven objects showing a range from 0.026 to 0.26, implying that  
there is little sign of 
destruction of the Ly$\alpha$\ line.   All of the properties of the $z=6.5$ sample
appear to be very similar to those of Ly$\alpha$ emitters at lower
redshifts. 
\end{abstract}


\newpage


\section{Introduction}
\label{secintro}

At all redshifts from $z=0$ to $z=7$ a fraction of galaxies can be detected
in the Lyman alpha line of hydrogen (e.g. Cowie and Hu 1998; Deharveng et al.\ 2008;
Hu et al.\ 2010; Ouchi et al.\ 2010; Vanzella et al.\ 2010).  
The Ly$\alpha$ line is easily detected in narrow-band imaging
or blind spectroscopic surveys, and 
Ly$\alpha$\ emission-line searches have been widely
used to find high redshift galaxies. In addition, for the
highest redshift galaxies, this line
is the only spectroscopic signature that can be used
to confirm the redshift of a galaxy selected on the basis
of color properties. However,
Ly$\alpha$ is a difficult line to interpret.
Because the line is resonantly scattered
by neutral hydrogen, determining its escape path and
therefore its dust destruction is an extremely complex problem
both theoretically (e.g., Neufeld 1991; Finkelstein et al.\ 2007)
and observationally (e.g., Kunth et al.\ 2003; Schaerer \& Verhamme 2008,
\"{O}stlin et al.\ 2009). It is therefore critical to determine
the ultraviolet continuum properties of the Ly$\alpha$\ emitting
galaxies (LAEs) in order
to understand how to interpret them in terms of physical
quantities such as the star-formation rate and to relate
them to UV-continuum selected samples at the same redshift
(e.g., Bouwens et al.\  2010).

 Extending this measurement to the highest redshifts possible
is of considerable interest because the LAEs may be used as
probes of the structure of the intergalactic gas at $z>6$
(e.g., Dijkstra \& Wyithe 2010, Tilvi et al.\ 2010, Dayal et al.\ 2011, Laursen et al.\ 2011).
Since Ly$\alpha$\ photons resonantly scatter off of neutral hydrogen, 
the Ly$\alpha$\ emission from galaxies scatters from the line of sight 
as it propagates through the intergalactic medium (e.g. Loeb \& Rybicki 1999; Zheng et al.\ 2010) 
potentially reducing the strength of the observed Ly$\alpha$.  
Thus the frequency of Ly$\alpha$\ emitters relative
to the UV continuum-selected galaxy population is potentially a 
function of the degree of ionization of the intergalactic medium 
and could be used to probe the epoch of reionization of the universe,
at redshift $z > 6$ (e.g. Stark et al.\ 2010, 2011). 

Here we extend the measurement of the UV-continuum properties
of the LAEs to $z=6.5$, the highest
redshift where substantial samples of these objects exist,
using very deep near-infrared continuum observations 
of a uniform sample of $z=6.5$\ LAEs obtained with the WFC3 camera on HST.
These are the first observations deep enough to measure individual 
continuum magnitudes at this redshift for all the objects in this population 
and we find that their UV-continuum properties and the fraction
of Ly$\alpha$ emitters relative to UV-continuum selected samples
are very similar to that seen at lower redshift. The observations
show no indication that the onset of reionization is at $z=6.5$. 

\section{Observations}
\label{secobs}

We obtained $1.4~\micron$\ (rest frame $1850\rm\AA$) continuum detections for a sample of
7 Ly$\alpha$\ emitters (LAEs) at $z=6.5$\  in a 600~arcmin$^2$\ 
region centered on the massive $z=0.3$ galaxy cluster Abell 370.
The sample was drawn from Hu et al.'s (2010) 
Ly$\alpha$ spectroscopic atlas of 88 $z=5.7$ and 30 $z=6.5$ galaxies. 
The Hu et al. Ly$\alpha$\ emitters were chosen based on
their narrow-band excess and their broad-band colors in
very deep images obtained with the Suprime camera on the Subaru~8m telescope.
They were then spectroscopically confirmed using the DEIMOS
spectrograph on the Keck~10m telescope. They form a sample
of objects with Ly$\alpha$ rest frame equivalent widths above
20\AA\ and log(L(L$\alpha$))$=42.5-43$ erg s$^{-1}$ in the selected redshift interval.
 These objects have narrow Ly$\alpha$\ emission lines 
(velocity widths 150 -- 300~${\rm km\ s}^{-1}$) and, 
as we shall show below, while they are small, they are marginally spatially extended in the present observations.  We therefore assume 
that they are powered by star formation and not by an active galactic nucleus.

We used the WFC3 camera on $HST$\ with the F140W
filter to obtain $1.4\micron$ observations
of all of the $z=6.5$ objects in the A370 field of that survey.  
Each of the $z=6.5$\ galaxies  was observed for 2 orbits using the
F140W filter of the WFC3 camera on HST (Mackenty et al.\ 2010). A four point dither pattern
was used and repeated 3 times, giving a total exposure time of 4235s
for each field. We used the combined drizzled images 
together with the calibration produced by the HST pipeline,
but then removed a further background determined from the
median average of the fields. 
The measured FWHM for faint stars in the images
is 0.22 arcseconds.
As we shall discuss below, the galaxies are small so,
given the high spatial resolution of the HST images, we 
used $0.8^{\prime\prime}$\ diameter magnitudes and
corrected these to approximate total magnitudes using
an offset of 0.15 magnitudes computed from the enclosed energy curves of the instrument. We measured the noise
in the images by measuring the fluxes in random blank sky positions
and found a $1~\sigma$\ noise level of 28.75 for the corrected
$0.8^{\prime\prime}$\ diameter magnitudes.


\begin{figure}[h]
\begin{center}
\includegraphics[angle=90,width=3.0in]{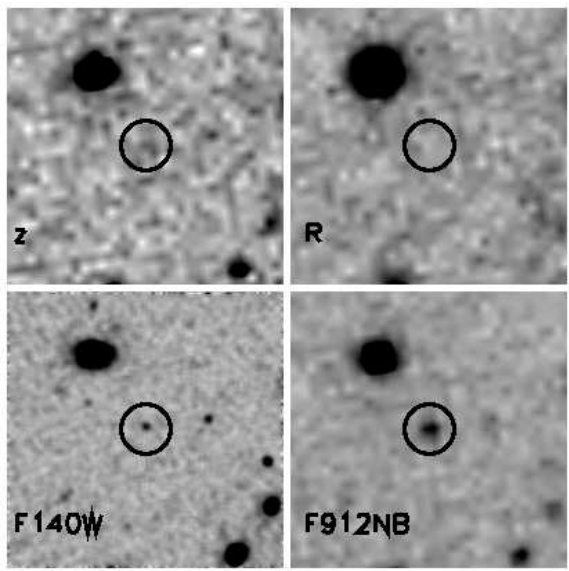}
\figurenum{1}
\caption{Comparison of the HST F140W observation
for a typical object (HC023939-013451)
with the ground based data from Hu et al.\ (2010). The 
lower left panel shows the F140W
image, the lower right the 9120\AA\ narrow band image,
the upper left the z band image and the upper right
the R band image. The circle marking the LAE is one arcsecond in
radius.
\label{typical_image}
} 
\end{center}
\end{figure}

All of the galaxies in the sample were detected; 
a typical example is shown in Figure~\ref{typical_image}.
The faintness of the image, even in a very deep
z band exposure such as that shown  ($5\sigma \sim 25.4$\ AB), illustrates the difficulty
of making accurate continuum measuruments from the ground.
The observed continuum magnitudes and other quantities 
are listed in Table~\ref{tab1}.  
The first detected $z=6.5$ galaxy, HCM6A (Hu et al.\ 2002), is
a member of the sample and since this galaxy lies 
in the strong-lensing region of A370,
for this object we also give the values corrected
for lensing amplification (quantities in parentheses 
in Table~\ref{tab1}).  All of the remaining galaxies lie far outside the strong lensing region and have negligible amplification.


\begin{deluxetable*}{lrlcccc}
\tabletypesize{\footnotesize}
\tablecaption{Galaxies in the sample\label{tab1}}
\tablenum{1}
\tablehead{\colhead{Name}  &
\colhead{RA(2000)} &
\colhead{Dec(2000)} &
\colhead{Redshift} &
\colhead{$\log L(\rm Ly\alpha$)} &
\colhead{mag$^a$} &
\colhead{Rest EW$^b$} \\
\colhead{} &
\colhead{} &
\colhead{} &
\colhead{} &
\colhead{(erg s$^{-1}$)} &
\colhead{(AB)} &
\colhead{(\AA)} 
}
\startdata
HC023927-013523  &     39.863293  &    -1.589833   & 6.450 & 42.73 & $26.68 \pm 0.16$ & $64 \pm 9$ \\
HC023939-013432  &     39.914417  &    -1.575667   & 6.549 & 42.76 & $26.03 \pm 0.09$ & $36 \pm 3$\\
HC023939-013451  &     39.914791  &    -1.581028   & 6.531 & 42.90 & $26.73 \pm 0.17$ & $97 \pm 15$ \\
HC023949-013121  &     39.956711  &    -1.522667   & 6.564 & 42.99 & $28.00^{+0.70}_{-0.42}$ & $357^{+329}_{-116}$ \\
HCM6A            &     39.978001  &    -1.559111   & 6.559 & 43.41(42.75$^c$) & $24.62(26.27^c)\pm 0.02$ & $44 \pm 1$ \\
HC024001-014100  &     40.007500  &    -1.683388   & 6.545 & 42.72 & $27.84^{+0.61}_{-0.39}$ & $175^{+134}_{-53}$ \\
HC024004-012252  &     40.017708  &    -1.381278   & 6.502 & 42.62 & $26.37 \pm 0.12$ & $37 \pm 4$ \\
\cutinhead{}
HC123607+620838 & 189.033004  &  62.14394  & 5.640 & 42.95 & $26.36\pm 0.12$ & $86\pm 10$ \cr
HC123558+621017 & 189.045471  &  62.17144  & 5.672 & 42.58 & $26.54\pm 0.18$ & $ 42\pm 8$ \cr
HC123613+620748 & 189.056106  &  62.12994  & 5.635 & 42.95 & $25.99\pm 0.15$ & $60\pm 9$ \cr
HC123651+621936 & 189.215240  &  62.32683  & 5.675 & 42.96 & $26.89\pm 0.24$ & $140\pm 34$ \cr
HC123652+622152 & 189.216751  &  62.36460  & 5.689 & 42.92 & $26.37\pm 0.22$ & $79\pm 18$ \cr
HC123717+621759 & 189.324677  &  62.29974  & 5.663 & 42.91 & $26.52\pm 0.15$ & $89\pm 13$ 
\enddata

\tablecomments{($a$)~Continuum magnitude in the AB system at $1.4~\mu m$\ ($z = 6.5$) 
and $0.91~\mu m$\ ($z = 5.7$).  
($b$)~Rest-frame equivalent width of the Ly$\alpha$\ line calculated assuming a 
flat-f$_{\nu}$\ spectrum for the continuum at wavelengths between Ly$\alpha$ and $\sim1850\rm\AA$\  
(rest-frame; $z = 6.5$) or $\sim1350\rm\AA$\  (rest-frame; $z = 5.7$).
($c$)~Values corrected for lensing amplification.}

\end{deluxetable*}

For a comparison sample at $z=5.7$\ chosen in the same way 
and with comparably deep continuum magnitudes,
we took Hu et al.'s sample of six $z=5.7$ spectroscopically
confirmed objects in the GOODS-N field, using
published  F850LP data obtained with the ACS on the
$HST$\  (Giavalisco et al.\ 2004) to define our continuum magnitudes.
The $z = 5.7$\ magnitudes and errors are the auto mags from 
the catalog of Giavalisco et al.\ (2004);  they also 
provide a good representation of the total magnitude.
The rest-frame wavelength corresponding to the pivot wavelength of this filter is 1350\AA.
While the F140W filter has no response at the Ly$\alpha$\ line,
the F850LP filter has 10\% of peak response at $8140~\rm\AA$\ 
(Ly$\alpha$\ for $z=5.7$).  However, the emission-line contribution
to the continuum fluxes is small ($<20\%$) even for 
the highest equivalent width (EW $\approx$\ 140\AA)  $z=5.7$\ galaxies.
The properties of these objects are also listed in Table~\ref{tab1}.

\section{Discussion}
\label{secdiscuss}

\begin{figure}[h]
\begin{center}
\includegraphics[angle=0,width=3.4in]{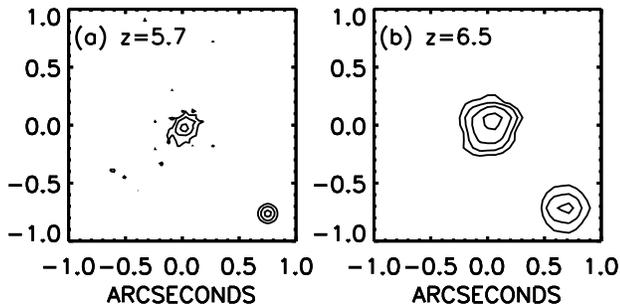}
\figurenum{2}
\vskip -1.0cm
\caption{Stacked rest-frame near-UV images for (panel a) the
$z=5.7$\ Ly$\alpha$-selected galaxies and (panel b) the subset of
isolated and unlensed $z=6.5$\ Ly$\alpha$-selected galaxies.
The point spread function measured from faint stars
in the field is shown in the lower right corner. In
both cases the images are resolved, 
though the effect is much more easily seen in
the higher resolution ACS images of the $z=5.7$
galaxies. 
\label{stack}
} 
\end{center}
\end{figure}

The LAE galaxies  at both $z=5.7$\ and $z=6.5$\  are  
extremely compact (Figure~\ref{stack}).  For each redshift sample we formed a stack by summing all the images, centering on the peak flux.  For the $z = 6.5$\ objects we excluded HCM6A, which is lensed, and HC023949$-$013121, which lies close to a bright galaxy.  
A comparison of the 
stacked continuum images of the two samples with the point 
spread function (PSF) determined from faint spectroscopically 
confirmed stars in
the two fields, shows that images at both redshifts are
extended:  the deconvolved FWHM is $0.08\arcsec$\ at $z=5.7$
and $0.14\arcsec$\ at $z=6.5$, corresponding to sizes between
0.5 and 1 kpc. This is very similar to values measured 
for Ly$\alpha$\ emitters (LAEs) at $z=3.1$\ (Bond et al.\ 2009, 2010), where 
both the UV continuum and the Ly$\alpha$ emission are seen to be sub-kpc. 
The sizes are smaller than those
found by Taniguchi et al.\ (2009) at $z=5.7$ but
the Taniguchi measurements were made in a bandpass
containing the Ly$\alpha$ line and it is possible that this 
could be increasing the measured size in their sample if 
the Ly$\alpha$ emission is more
extended.  The UV size measurements provide 
a reference for future measurements of the Ly$\alpha$\ 
sizes, which will allow us to measure the extent 
of the Ly$\alpha$\ scattering region in galaxies.  Such measurements 
are difficult but may be possible with ground-based adaptive optics 
or space-based narrow-band measurements.


\begin{figure}[h]
\begin{center}
\includegraphics[angle=0,width=3.5in]{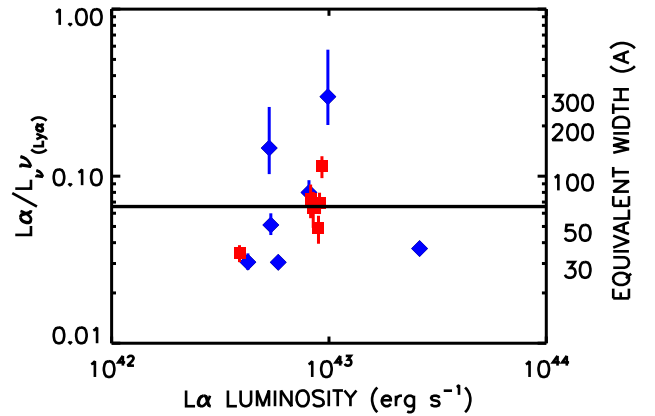}
\figurenum{3}
\caption{Fraction of light coming out in the Ly$\alpha$ line
versus that coming out in UV continuum redward of Ly$\alpha$. 
(The latter is calculated assuming a flat-$f_\nu$ continuum).
This fraction is linearly related to the Ly$\alpha$ equivalent width
which is given on the right hand vertical axis of the figure.
Objects at $z=6.5$ are shown with 
blue diamonds and objects
at $z=5.7$ with 
red squares. 
The error bars are $\pm 1~\sigma$.  
\label{lplot}
}
\end{center}
\end{figure}

Figure~\ref{lplot} shows our primary result.  For both samples 
we plot the ratio of Ly$\alpha$ luminosity to 
continuum luminosity redward of Ly$\alpha$ (or equivalently 
the rest-frame equivalent width of the Ly$\alpha$ line, to 
which the ratio is approximately proportional [Dijkstra \& Westra 2010]) 
as a function of Ly$\alpha$ luminosity.  Following 
Stark et al.\ (2010), we have assumed $f_{\nu} \sim \nu^0$\ 
in computing the equivalent width.  Adopting $f_{\nu} \sim \nu^{-0.8}$\ 
as suggested in Taniguchi et al.\ (2009) would increase 
the equivalent widths by a multiplicative factor of 1.4 
for the $z=6.5$\ systems and 1.1 for the $z=5.7$\ systems.  However,
we note that we focus less on the Ly$\alpha$\ equivalent width than on the UV continuum luminosity, which
is less sensitive to the assumed $f_{\nu}$  since the wavelength difference is smaller.  Here too we assume $f_{\nu} \sim \nu^0$\ in comparing the two redshifts.    
The $z=5.7$ sample has identical median and mean values
of 0.07 for the Ly$\alpha$\ to UV continuum luminosity ratio, while the $z = 6.5$\ sample has a median value of 0.05 and a mean value of 0.08.
Both are very similar to that expected
from a galaxy undergoing constant star formation. 
For a Salpeter initial mass function, the latter value 
is $\approx 0.065$\ almost irrespective of metallicity 
(Charlot \& Fall 1993; Schaerer \& Verhamme 2008).  This value is 
shown as the solid black line in Figure~\ref{lplot}.
There is little sign of evolution
with redshift, therefore, and, for this sample,
little sign of destruction of the line relative to the continuum.  The observed values catter above and below the average values by multiplicative values of roughly three.  The high values may correspond to younger age, top-heavy IMF or particular escape geometry (e.g. Finkelstein et al.\ 2007).

%
%

The high value of the Ly$\alpha$\  to UV continuum ratio is likely to be, at least in part, a selection
effect: if there is a range in the amount of 
destruction of Ly$\alpha$\ relative to the 
continuum, then the most luminous Ly$\alpha$
emitters will be those with the least destruction, which 
may then have near-standard ratios.  In particular, objects 
without Ly$\alpha$\ detections at the rest-frame EW limit 
of $20~\rm\AA$\ should not be analyzed with this result, which 
holds only for objects in the LAE samples.  However, 
the result does imply that many objects with 
little Ly$\alpha$\ destruction do exist and that, 
for these luminous Ly$\alpha$-selected objects, we can convert 
between Ly$\alpha$ and UV continuum luminosities.  At first sight this may seem inconsistent with the blue asymmetric profiles seen in these Ly$\alpha$\ emitters (Hu et al.\ 2010).  However, these profiles may be intrinsic to the galaxy and produced by scattering in outflows (Verhamme et al.\ 2006) rather than being a consequence of Ly$\alpha$\ scattering in the surrounding intergalactic gas.

Our measured ratio of Ly$\alpha$\ luminosity to UV continuum 
luminosity can be combined with the `conventional'
Kennicutt (1998) relation between the star formation rate 
and the UV luminosity to obtain 
$\log {\rm SFR(M_\sun yr^{-1}}) = \log L({\rm Ly}\alpha)-42.03$,
where $L({\rm Ly}\alpha)$ is in erg s$^{-1}$\ and, again, the value is for a Salpeter initial mass function. The effects
of metallicity may change this relation by as much
as 0.2 dex  (Schaerer \& Verhamme 2008) and it is also 
dependent on the exact star formation history.
However, it can be seen from the Ly$\alpha$\ luminosities in
Table~\ref{tab1} that typical 
star formation rates, with no extinction correction,
are in the range of 5--10 M$_\sun$ yr$^{-1}$, where we have used the relationship above in computing the SFRs.  
The UV continuum-selected galaxies without detected Ly$\alpha$\ may 
have much more substantial Ly$\alpha$\  destruction and for these 
the conversion rate from Ly$\alpha$\ luminosity to SFR may be
larger than the relation given above.

%
%
\begin{figure}[h]
\begin{center}
\includegraphics[angle=0,width=3.5in]{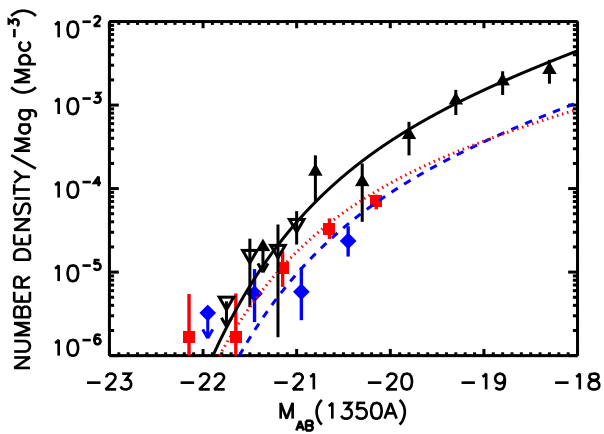}
\figurenum{4}
\caption{Ly$\alpha$-selected 1350\AA~continuum luminosity functions
at $z=5.7$ (red squares) and $z=6.5$ (blue diamonds)
derived from the Hu et al.\ (2010) Ly$\alpha$ sample
at these redshifts. Errors are $\pm 1\sigma$
and upper limits are $1\sigma$\ and shown with a downward
pointing arrow. The black filled triangles show the $z=6.8$
luminosity function derived in Bouwens et al.\ (2010) 
and the open downward pointing triangles that of
Ouchi et al.\ (2009) while the solid black curve
shows the Bouwens et al.\ (2010) maximum likelihood 
Schechter-function fit to the combination of these two samples.
The blue dashed curve
is this fit renormalized by a factor of 0.24 in number density
to match the Ly$\alpha$-selected $z=6.5$ continuum
luminosity function. The red dotted
curve is the  $z=5.9$ color-selected
continuum luminosity function from Bouwens et al.\ (2010) 
renormalized by a factor of 0.18 in number density to match the $z=5.7$
Ly$\alpha$-selected continuum luminosity function.
\label{mab_funs}
} 
\end{center}
\end{figure}


We can also now test whether the Ly$\alpha$\ emitters are comparable in continuum luminosity to samples obtained using continuum color-selection.
From the full LAE sample of Hu et al.\ (2010) we constructed continuum luminosity functions of the
Ly$\alpha$-selected sample at both $z=5.7$ and $z=6.5$, using the accessible volumes computed for each object in Hu et al.  We converted 
Ly$\alpha$ luminosity  to a UV continuum luminosity at 1350\AA\ (rest-frame) using the 
constant star-formation value of 0.065 
for the ratio of Ly$\alpha$\ luminosity to UV continuum light
redward of Ly$\alpha$, and assuming the continuua are roughly
constant in $f_{\nu}$ from 1350\AA\ to 1850\AA.  These luminosity functions are shown in 
Figure~\ref{mab_funs}, where we compare them with continuum luminosity functions determined from color-selected samples (Ouchi et al.\ 2009; Bouwens et al.\ 2010).   
The Ly$\alpha$-selected objects have high continuum luminosities (or equivalently star formation rates), comparable to the brightest continuum-selected objects at these redshifts.  Our UV continuum luminosity function is lower by a factor of 5 than those derived by Shimasaku et al.\ (2006) and Kashikawa et al.\ (2006) whose results would imply that all UV continuum-selected objects at this redshift would also be LAEs.  Both of these results may be due in part to the extreme difficulty of measuring accurate continuum magnitudes from ground-based data, particularly when the Ly$\alpha$\ lines lie within the continuum bandpass. 
However, the difference is also due in part to the high Ly$\alpha$\ 
luminosity functions calculated in these papers, which are about 
a factor of two above that of Hu et al. (2010).  Hu et al.\ (2010) 
argue that the difference between their Ly$\alpha$\ LF and 
the previous determinations is a consequence of using a fully spectroscopically confirmed sample which eliminates some of the photometrically selected objects but Kashikawa et al.\ (2011) argue against this interpretation.  It is unclear therefore which is the better Ly$\alpha$\ LF to choose, but we note that using the higher Kashikawa et al.\ (2011) Ly$\alpha$\ LF would result in a very large fraction of UV continuum-selected galaxies having Ly$\alpha$\ emission at near the case B ratio.  This would be a rather surprising result, given that we expect at least some destruction owing to scattering in the IGM.

We can also use the luminosity functions to estimate the fraction
of UV continuum sources that have strong Ly$\alpha$
emission. This fraction
rises rapidly at low redshifts (from 5\% at $z=0$\ to $\sim 20$\% at $z=2$; Cowie, Barger \& Hu 2010)  but 
above $z=2$\ it has been less clear whether
it continues to increase or drops again (Stark et al.\ 2010, 2011; Zucca et al.\ 2009; Hayes et al.\ 2010).  
Above a UV continuum magnitude of $-20.2$\ the Ly$\alpha$ sample
comprises $24\pm6\%$ of the color-selected continuum sample at $z=6.5$.
For $z=5.7$ the corresponding number is $18\pm2\%$ above
$-19.9$, where the errors are formal statistical errors. These fractions are very similar to those found
at $z=3$ by Shapley et al.\ (2003) for sources with
similar rest-frame Ly$\alpha$ equivalent widths
(above 20\AA), which would suggest very little high-redshift evolution.
It would represent a drop from the values at $z = 5.6$ - 6.1 found by Stark et al.\ (2011) who argue that nearly all continuum-selected objects at these redshifts are LAEs with rest-frame EW $> 20\rm\AA$.
However, we caution that the present numbers have substantial systematic uncertainties. The uncertainty in the UV to Ly$\alpha$\ conversion ratio can result in significant changes:  thus if the true conversion was 0.03 (0.13), the second lowest (second highest) value observed in the sample, the normalization would rise (fall) to approximately 0.6 (0.1).  
In addition, as we have discussed above, other measurements
of the Ly$\alpha$ luminosity function at these redshifts (Kashikawa et al.\ 2006, 2011; Ouchi et al.\ 2008, 2010) 
have been a factor of two higher than Hu
et al.'s and using these luminosity functions would increase the LAE fraction by this amount.  It is also possible that the UV continuum
samples we are comparing with may still contain spurious objects,
including red stars and lower redshift red galaxies, which would also increase the LAE fraction, and that there may be substantial uncertainties owing to cosmic variance in these continuum luminosity functions. 

  The present results show that LAEs at $z=6.5$ have remarkably
similar properties to those at lower redshifts. They are small,
have moderately high star formation rates, and have substantially
complete escape of
the Ly$\alpha$\ photons. Other properties such as the shape
and widths of the Ly$\alpha$ lines also appear invariant 
(e.g. Ouchi et al.\ 2010, Hu et al.\ 2010). The results
also show that the continuum luminosities are comparable to the
brightest values seen in color-selected continuum samples, 
suggesting that even the strongest star-forming galaxies
at these redshifts may show Ly$\alpha$ in emission.
We also find that the LAEs are a significant fraction of the UV-continuum
selected objects at these redshifts, similar, within the systematic uncertainties, to values seen at lower redshifts. The observations show
no sign that any effects associated with reionization are
being seen at $z=6.5$. 


\acknowledgements
We gratefully acknowledge support from NSF
grants AST-0709356 (L.~L.~C.), AST-0687850 (E.~M.~H.), and AST-0607871 (A.~S.) and 
from Hubble Space Telescope HST-GO-11108.01-A (E.~M.~H.)

\newpage
\clearpage

\end{document}